\documentclass[aps,prd,twocolumn,groupedaddress,floatfix]{revtex4}

\newif\ifpdf
\ifx\pdfoutput\undefined
   \pdffalse
\else
   \pdfoutput=1
   \pdftrue
\fi

\usepackage{latexsym}
\ifpdf
   \usepackage{thumbpdf}
   \usepackage[pdftex,dvipsnames,usenames]{color}
   \usepackage[pdftex,
       colorlinks=true,
       urlcolor=blue,       
       filecolor=green,     
       linkcolor=red,       
       pdftitle={The Nonlinear Sigma Model With Distributed Adaptive Mesh Refinement},
       pdfauthor={Steven L. Liebling},
       pdfsubject={Distributed AMR},
       pdfkeywords={amr mpi},
       pagebackref,
       pdfpagemode=None,
       bookmarksopen=false]{hyperref}
   \usepackage[pdftex]{graphicx}
\else
   \usepackage{graphicx}
   \usepackage{color}
   \usepackage{epsfig}
   \def\href#1#2{#2}

\fi

\def\lb{\left(}
\def\rb{\right)}
\def\be {\begin{equation}}
\def\ee {\end{equation}  }
\def\beq{\begin{eqnarray}}
\def\eeq{\end{eqnarray}  }
\def\bi {\begin{itemize} }

\def\ei {\end{itemize}   }
\def\RE {I\kern-6pt R    }
\def\Z  {Z\kern-13pt Z   }
\def\be {\begin{equation}}
\def\ee {\end{equation}  }
\def\beq{\begin{eqnarray}}
\def\eeq{\end{eqnarray}  }

\def\eeq{\end{eqnarray}  }

\begin{document}

\title{The Nonlinear Sigma Model With Distributed Adaptive Mesh Refinement}

\author{Steven L. Liebling}
    \affiliation{Southampton College, Long Island University,
                 Southampton, NY 11968}

\date{\today}

\begin{abstract}

An adaptive mesh refinement (AMR) scheme is implemented in a distributed
environment using Message Passing Interface (MPI) to find solutions to
the nonlinear sigma model.
Previous work studied behavior similar to
black hole critical phenomena at the threshold for singularity formation
in this flat space model. This work is  a follow-up describing extensions
to distribute the grid hierarchy and 
presenting tests
showing the correctness of the model.
\end{abstract}


\maketitle

The problem of modeling collisions of two black holes in general
situations poses a number of very difficult problems.
One among these is finding enough computational resources
to adequately resolve the physics in a reasonable amount of time.
Adaptive mesh refinement (AMR), whereby fine grids are added
where and when needed, has proved successful at
enabling a tremendous dynamic range in resolution. However, even
with the efficiencies provided by AMR, single  machines
often fail to provide adequate memory or speed. A logical next step
is then to distribute the AMR problem across a cluster of processors.

Perhaps the most well known use of AMR in relativity is Choptuik's 
implementation of gravitational collapse in spherical
symmetry~\cite{choptuik93}.
Studying the threshold of black hole formation required a large
range in resolution to resolve the unique behavior he discovered in so called
black hole critical phenomena.
Other efforts with AMR include modeling
a single black hole~\cite{brugmann,centrella2},
perturbations of a black hole~\cite{wild},
binary black hole initial data~\cite{diener},
gravitational waves~\cite{centrella},
various cosmological models~\cite{hernthesis},
a scalar field using characteristic coordinates~\cite{pretorius03},
fixed mesh refinement~\cite{hawley},
orbiting black holes~\cite{tichy},
and
axisymmetric gravitational collapse~\cite{graxi1,graxi2}.

I describe an implementation of distributed AMR in a model problem 
much simpler than gravitational collapse, but which
nevertheless poses an interesting computational problem. In particular,
I find solutions to the nonlinear sigma model using a parallel 
implementation of AMR as a step towards extending the same computational
infrastructure to solve the gravitational field equations.

Previously~\cite{myamr}, solutions to this model were found using
serial AMR (a single machine) which showed a type of threshold behavior
similar to black hole critical behavior~\cite{choptuik93}. Families
of initial data parameterized by some parameter $p$ demonstrated two
types of behavior. For small $p$, energy dispersed to infinity
(the so-called {\em sub-critical} regime). For large $p$, evolutions
suggest that a singularity forms at finite time ({\em super-critical}).
Tuning between these two states (dispersal and singularity formation),
one finds the {\em critical} regime $p\approx p^*$ in which solutions
approach a unique
solution independent of the family with which one begins.
This solution is the {\em critical solution} and demonstrates
self-similarity.

This paper serves as a follow-up to~\cite{myamr}.
Presented first are the model equations. I then discuss
details about the implementation of the
model and the distributed AMR algorithm.
Tests of the numerical code follow along with
a discussion of the parallel performance of the code.

{\em \bf The model:}~~The nonlinear sigma model provides an interesting toy model with which
to explore distributed AMR. Choosing a generalized hedgehog ansatz~\cite{myamr},
the dynamics of the model reduce to 
a single equation of motion for a scalar field $\chi(x,y,z,t)$
\be
\ddot \chi =
                   \chi_{,xx}
                  + \chi_{,yy}
                  + \chi_{,zz}
             -\frac{ \sin 2\chi}{r^2},
\label{eq:eom}
\ee
where commas indicate partial differentiation with respect to subscripted
coordinates, an overdot denotes $\partial / \partial t$, and $r \equiv
\sqrt{x^2+y^2+z^2}$. The equation of motion~(\ref{eq:eom}) implies the
regularity condition $\chi(0,0,0,t)= 0$ which is enforced by the evolution
procedure. The energy density associated with this system is given by
\be 
\rho = \frac{1}{2} \left[
                                     \lb \dot \chi \rb^2
                                   + \lb \chi_{,x} \rb^2
                                   + \lb \chi_{,y} \rb^2
                                   + \lb \chi_{,z} \rb^2
                                   \right]
                + \frac{\sin^2 \chi}{r^2},
\ee
while the angular momentum densities are given by~\cite{ryder}
\be
M^{\mu \nu} = \int d^3 x \left( T^{0\mu} x^\nu - T^{0\nu} x^\mu \right),
\ee
where the $z$-component of the angular momentum, for example, is
\be
J_z = \int d^3 x~M^{xy} = \int d^3 x ~\dot \chi \left( y \chi_{,x} - x \chi_{,y} \right).
\ee
Initial data for $\chi$ at $t=0$ is set to a generalized Gaussian pulse, and then
evolved using the equation of motion~(\ref{eq:eom}) with second-order finite differences
in an iterative Crank-Nicholson scheme.

{\em \bf Implementation:}~~
Certain solutions of this model require a tremendous
dynamic range achievable only with AMR.
The method implemented here follows that of
Berger and Oliger~\cite{berger} though with certain simplifications. The general strategy
is that during the course of an evolution, the dynamics are monitored such that when more
resolution is needed, finer sub-grids are created in the regions which demand it. Concurrently,
when the dynamics no longer dictate the existence of fine grids, they are removed.
The description which follows is intended only to present the choices and simplifications
specific to this implementation and therefore assumes some
familiarity with the general algorithm as described in~\cite{berger}.

Consider a grid with uniform grid spacing $h = \Delta x = \Delta y = \Delta z$.
Some refinement criterion  is chosen to determine which points on the grid require
increased resolution, a process called {\em flagging}.
In this model, points
for which the energy density obeys $ h^2 \rho > \epsilon$, for some threshold value $\epsilon$,
are flagged. This condition compares the size of the squared derivatives, $\rho$, to the resolution
squared, $1/h^2$.
Other refinement criteria are widely used including estimates of truncation
error, but this choice works well for this model. 

Once points are flagged, bounding boxes are established
 by {\em clustering} those points 
into rectangular regions where sub-grids of finer resolution will be created. 
One simplification assumed in this code is that grids are completely nested within their parent which means that grids have unique parents.
This restricts the types of clustering allowable, and here a very simple method is used.
A bounding box is found for each disconnected set of  flagged  points.
In addition, a buffer region around all flagged points is included.
Points within two grid points of a flagged point are themselves flagged.
This buffering helps ensure
that moving features needing refinement stay within refined regions at least until the next refinement
occurs.

Each grid's resolution is an integral multiple of its parent's, called the {\em refinement factor}.
This implementation requires this factor to be even.
Newly created sub-grids are first initialized by linear interpolation in space from their
parent. Then
information from other grids at the same resolution ({\em siblings}),
including those destined for removal,
is used, where available, to initialize the grid. Around the boundaries of regions
initialized from siblings, points are linear interpolated
in space to smooth out irregularities between the data from the parent and from its siblings.

Advancement in time of a sub-grid follows that of coarser grids. In particular, once its
parent takes a time step, boundary values for advanced time steps of the sub-grid are
interpolated in time and space from the parent. The sub-grid then takes a number of steps
equal to the refinement factor until it is time aligned with its parent. At that point,
the parent grid is injected via half-weighted
restriction from values in the sub-grid.

\begin{figure}
\centerline{\includegraphics[clip=true, width=16cm]{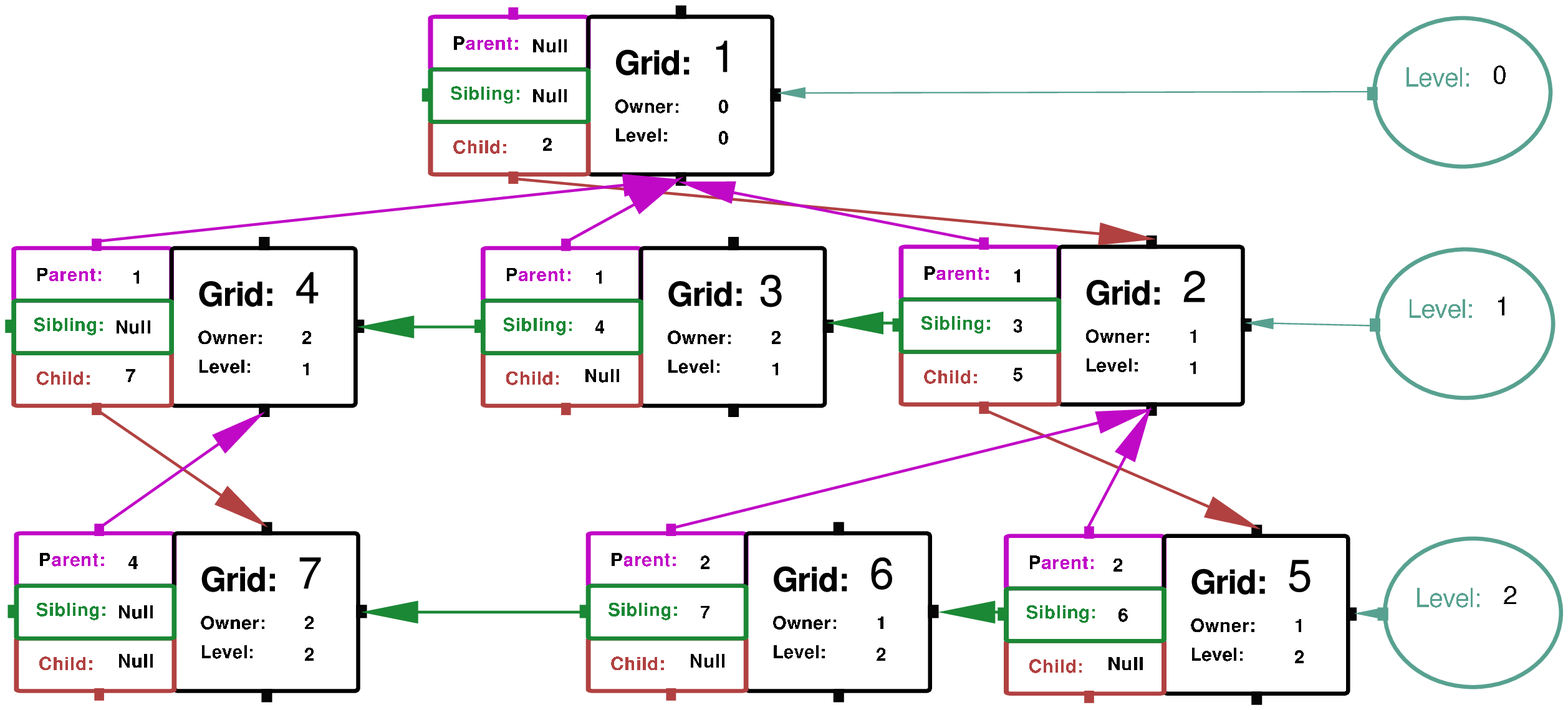}}
\vspace{-3mm}
\caption{\label{fig:tree}Diagram detailing the structure of the stored grid
hierarchy. Shown is a simple example of a three-level tree with grids represented
as rectangles. Each grid has a unique grid number and also stores its {\em parent},
its next {\em sibling}, and a single {\em child}.
In instances in which a sibling/child/parent
does not exist, a null pointer is used. The hierarchy also stores the process id which
``owns'' the grid (and therefore stores the associated fields defined on that grid).
Levels, which consist of all (sub-)grids at a given resolution are represented in the
diagram as ellipses. Levels store only the beginning grid number in the hierarchy, with
subsequent grids obtained by traversing the sibling pointers until a null pointer is reached.
}
\end{figure}

The grid hierarchy can be considered as the union of levels, where a level consists
of all grids at a given resolution. Thus the coarse grid constitutes Level 0,
all its immediate children fill out Level 1, its grandchildren form Level 2, and so on.
An example of a grid hierarchy and how
it is stored is shown in Fig.~\ref{fig:tree}.

A number of strategies  exist for distributing an AMR code. Hoping to achieve
a reasonably simple implementation of what is inherently complex, the method 
adopted here
is to distribute grids in their entirety to different processors. An alternative would
be to take a piece of every grid and place it on the various processors. Here, the different processors
``own'' different grids and are responsible for their time stepping and for any output
to data files.  However, all processors maintain their own copy of the grid hierarchy (as shown in Fig.~\ref{fig:tree})
but do not store the fields that live on grids owned by others.
Such a scheme necessitates that various inter-grid communication traverses
the network, except in the cases where the involved grids are owned by the same processor.

One advantage to this strategy is that much of the code written for a single processor 
(the {\em serial code}) carries over directly. In fact, much of the inter-grid routines carry over
with only minimal changes. For example, to provide boundary conditions for an advanced time step of
a sub-grid, boundary values are interpolated in time from a parent grid. In the serial version,
such values could be calculated in place, whereas in the distributed version
coarse grid values are interpolated in time on the coarse grid's processor, and are then sent
to the fine grid's processor where the values are interpolated in space. Such a scheme minimizes
the amount data needed to be transmitted and also allows the same routines to be used for the serial
version just without transmitting the data.

These grids need to be stepped in coordination so that when boundary values are needed by a child
grid, the parent is advanced to the correct time before those values are computed.
This coordination is provided by having
a {\em master} processor which owns the coarse grid with all other processors
entering {\em slave} mode.  In this mode, they  continually
loop responding to commands passed from other processors over the network.

\begin{figure}
\centerline{\includegraphics[width=10cm]{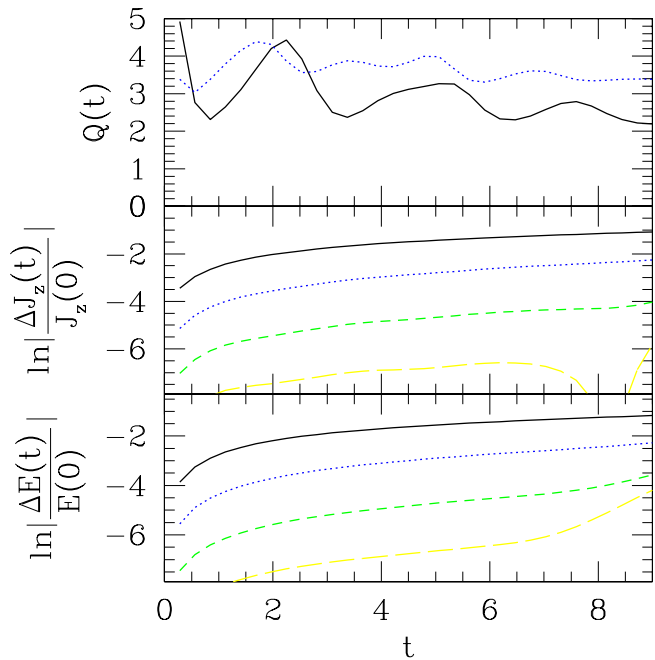}}
\caption{\label{fig:converge} Tests of a typical unigrid evolution with
numbers of points:
$\left(32 +1\right)^3$ (solid,                 black),
$\left(64 +1\right)^3$ (dot,                   blue),
$\left(128+1\right)^3$ (short dash,            green),
and
$\left(256+1\right)^3$ (long dash,             yellow).
The {\bf top frame} shows the 
convergence factor Q(t) as in Eq.~\ref{eq:cvf}. For low resolution, convergence is
poor, but as resolution increases, the convergence rapidly improves,
approaching the expected value $4$. The {\bf middle frame} shows
the loss of the $z$-component of the angular momentum as a function of time, where
$\Delta J_z(t) \equiv J_z(t) - J_z(0)$. As the resolution increases, the loss
decreases as would be expected for a convergent code. The {\bf bottom frame}
shows the loss of energy as a function of time. As with the angular momentum,
the loss converges toward zero. The dimensionless ratio of
the angular momentum to the energy squared is
$J/E^2 = 0.0072$.
}
\end{figure}

\begin{figure}
\centerline{\includegraphics[width=10cm]{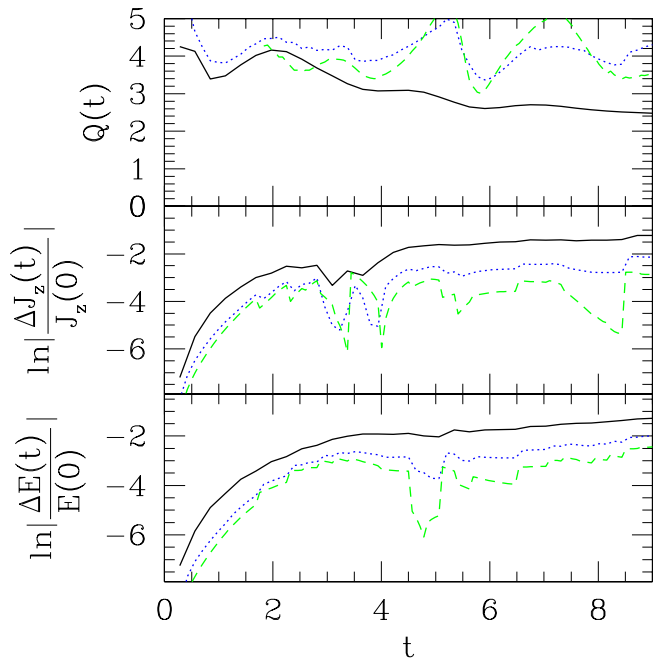}}
\caption{\label{fig:convergeAMR} Tests of an AMR evolution. The initial data is equivalent to
that shown in Fig.~\ref{fig:converge}. The three runs use identical adaptive hierarchies modulo
a factor of $2$ in resolution.
The {\bf top frame} shows the convergence factors for the respective coarse grids (solid, black),
the Level 1 grids (dot, blue), and the Level 2 grids (short dash, green) for the times that they exist.
The {\bf middle} and {\bf bottom frames} show the change in angular momentum and energy, respectively
computed as an integral over the entire grid hierarchies.
The data correspond to runs with coarse grids of resolution:
$\left(32 +1\right)^3$ (solid,                 black),
$\left(64 +1\right)^3$ (dot,                   blue),
and
$\left(128+1\right)^3$ (short dash,            green).
}
\end{figure}

In the serial version, for each coarse grid time step, appropriate
actions are taken on entire levels.  A similar approach works for the distributed code. For example, say 
Level 3 has been advanced, and it is now time to update the boundary values on all grids on Level 4. Then
the master processor loops over all grids on Level 4 owned by other processors and sends a command to those owner
processors to update the boundaries of the appropriate grids. It then, proceeds by updating the boundary values
of all Level 4 grids it owns. Finally, it enters slave mode in which it listens in case it owns grids which
must provide such boundary values.  Allowing each processor more autonomy would likely
increase scaling performance but would add to the complexity.

Distribution of the computational problem achieves two goals.
The first is that the computation is sped up with respect to running
a single processor. The second is that the memory available increases
with the number of nodes. Ideally, both these scale close to linearly
with the number of nodes. In the case of speed, optimal scaling can
only be achieved if the processors are kept busy, else it is possible
for no speed increase to occur. This problem is called {\em load balancing}
and merits quite a bit of literature with a variety of schemes.

Here, such concerns are postponed for later study, and instead a nearly trivial
scheme is implemented. That scheme involves maintaining a table of
the workload on each processor. Every subgrid of the coarse grid
is assigned to the least loaded processor. For other subgrids, they
are assigned to the process which owns their parents unless there
is a processor with no workload. A number of inter-grid operations
occur between a parent and a child, and hence the scheme favors
putting children on the same processor as their parent.

The distributed code is sufficiently similar to the serial version that only
the distributed version is maintained. Of course, it can still be run on
a single processor,
and the incumbent overhead is minimal.

Another benefit to such a code is
that it can accomplish domain decomposition in which  the computational
domain can be partitioned into equal chunks and processed on different 
processors. Decomposition enables one to take advantage of distributed
resources without adaptive mesh refinement, and is carried out be defining the zeroth level to
consist of multiple grids which cover the entire domain. A future goal is to be able adapt upon this
decomposed coarse grid.

{\em \bf Tests:}~~
A very strong test of the code was presented in the first paper~\cite{myamr}
in which the obtained critical solution was compared with results from
a 1D code~\cite{Liebling:1999nn}. These solutions span an extraordinary range
of resolutions, and that they agree is strong evidence that the codes model
the same system.

More traditional tests also provide evidence for the fidelity of the evolutions
to the proper equations. In particular, one can examine the behavior of numerical
solutions as resolution is increased. Representing a numerical solution at
some resolution $h$ by $\tilde \chi_h$, we can define a convergence factor
\be
Q(t) = \frac{ |\tilde \chi_{4h} - \tilde \chi_{2h} |_2 }
            { |\tilde \chi_{2h} - \tilde \chi_{ h} |_2 }
\label{eq:cvf}
\ee
in terms of solutions found for three successive resolutions $h$, $2h$, and $4h$.
If the solutions converge to a unique solution, this factor will be greater than unity,
and for second-order convergence one expects a factor of four.

In a similar fashion, one expects numerical solutions to maintain approximately conserved
quantities. Monitoring changes in the computed energy and angular momentum, one expects
these changes to converge to zero as resolution increases. As shown in Fig.~\ref{fig:converge},
for typical initial data, the convergence factor approaches four while changes in energy and
angular momentum about the $z$-axis converge to zero.

A similar test is presented in Fig.~\ref{fig:convergeAMR} in which the same initial data was
tested using AMR. Here, with a coarse grid of $\left( 33 + 1 \right)^3$ points, an evolution
allowing for three levels of refinement is run which outputs a history of sub-grids created.
The resolution of the coarse grid is doubled and run again with refinements dictated by the
refinement history recorded in the previous run. In this fashion,
adaptive runs can be duplicated with the only change being a multiple of the base resolution.
These results indicate convergence and can be compared to the results in Fig.~\ref{fig:converge}.


The code has also shown to give the same results to within machine
precision when run on 
a varying number of processors.

{\em \bf Performance:}~~
As already mentioned, the efforts described here are aimed at achieving
better performance by utilizing multiple processors, but are not expected
to achieve optimal performance. With that said, it makes sense to examine
the performance as a function of the number of processors.

With the design calling for entire grids to be placed on processors (as opposed to 
breaking all grids into smaller grids), there are two regimes of expected
performance. If each Level consists of a single grid, then the grids are nested
within each other and the grid hierarchy is completely {\em vertical}.
In this regime, generally little speedup is expected. The reason is that
the finest grid will usually require the most work,
and hence all other processors will generally have to wait for the finest grid
to complete before proceeding. Because the other grids take much less work
than the finest grid, if one were to run such an evolution on a single processor,
it would take about as long as on many.

The other regime is where there are many grids at the same resolution with
about the same number of points. For this case, the grid hierarchy is
spread {\em horizontally}. For such a system, the processors can truly
act in parallel, and a significant speedup is expected.

\begin{figure}
\centerline{\includegraphics[clip=true,width=10cm]{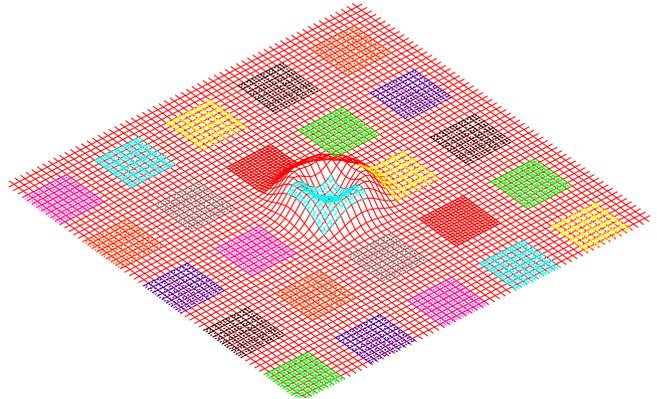}}
\vspace{-3mm}
\caption{\label{fig:125boxes} 2D slice of initial data with 125 sub-grids.
The data shows five equal-sized sub-grids in each dimension. The situation for
two, three, and four sub-grids per side ($8$, $27$, $64$ total sub-grids) is
similar. For so called ``static'' runs, no re-refinement was done, and for ``moving''
the sub-grids were successively moved one grid point with each refinement.
}
\end{figure}

As a first test, such a scenario (see Fig.~\ref{fig:125boxes}) is implemented by forcing the
code to create some number of equal sized grids as children of the coarsest
grid. We can define a speedup factor ${\cal S}$ in terms of the time $T_n$ it
takes for the code to complete on $n$ dual processor nodes as
\be
{\cal S} = \frac{T_1}{T_n}.
\label{eq:speedup}
\ee
$T_1$ is
the time taken on one dual processor node ({\em e.g.} two processors).

\begin{figure}
\centerline{\includegraphics[width=10cm]{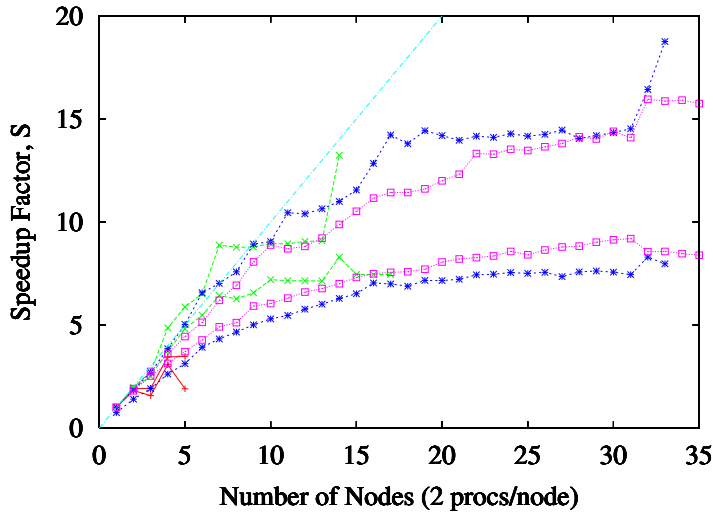}}
\caption{\label{fig:scaling}Scaling performance on an Itanium cluster ({\tt titan.ncsa.uiuc.edu} at NCSA).
Pairs of data are shown, one for so-called ``static'' and one for
''moving'' runs of a certain number of sub-grids. In general, the better performing
of the pair (the one on top) is the ``static.'' 
The various cases consist of: plus signs ($+$) for the $8$ sub-grid case;
                              crosses ($\times$) for the $27$ sub-grid case;
                              asterisks ($*$) for the $64$ sub-grid case;
                          and boxes ($\Box$) for the $125$ sub-grid case.
The dashed line indicates ideal, linear scale-up. The figure indicates significant
speed-up as long as an average of at least two grids exist per
process (in other words, the number of nodes is less than a quarter the number of sub-grids).
}
\end{figure}

These speedup factors are  shown in Fig.~\ref{fig:scaling}, for two
variations. The first is that these subgrids remain fixed while the second
has the subgrids moved a single grid point every few time steps. The latter
variation is to simulate grids which move and their incumbent network overhead.
The figure shows that the execution time is sped up but the improvement
generally saturates beginning when the number of processors exceeds half of
the total number of grids. In other words, good performance is predicated on
having many grids per processor.

A less contrived test would simply compare run times on various numbers of
processors. However, for this model, the essential dynamics consist of 
central collapse, and hence a typical, well resolved evolution would contain
a number of grids largely nested within each other in the so-called vertical regime.
As such, significant
speedup is not expected, but there is benefit in that distributing the problem
provides gains in memory storage. 


{\em \bf Conclusion:}~~
These results indicate the validity of the implementation of the nonlinear
sigma model, and in particular support the assertion that, within
the hedgehog ansatz but away from spherical symmetry, the 
critical solution between dispersion and singularity formation
is the same as that found under the assumption of spherical symmetry~\cite{Liebling:1999nn}.

This work serves as an initial step toward modeling general gravitational
collapse with distributed adaptive mesh refinement. 
Much work remains, including
the implementation of more sophisticated clustering and  load balancing.

{\em \bf Acknowledgments:}~~
This research was supported in part by NSF cooperative agreement ACI-9619020 through computing
resources provided by the National Partnership for Advanced Computational Infrastructure at
the University of Michigan Center for Advanced Computing and
by the National Computational Science Alliance under PHY030008N.
This research was also supported in part by NSF grants PHY03-25224 and PHY01-39980, by
Southampton College, part of Long Island University, and by NSF grant PHY99-07949 to the Kavli Institute of
Theoretical Physics at UCSB for whose hospitality I am thankful.

\bibliography{paper}

\begin{thebibliography}{16}
\expandafter\ifx\csname natexlab\endcsname\relax\def\natexlab#1{#1}\fi
\expandafter\ifx\csname bibnamefont\endcsname\relax
  \def\bibnamefont#1{#1}\fi
\expandafter\ifx\csname bibfnamefont\endcsname\relax
  \def\bibfnamefont#1{#1}\fi
\expandafter\ifx\csname citenamefont\endcsname\relax
  \def\citenamefont#1{#1}\fi
\expandafter\ifx\csname url\endcsname\relax
  \def\url#1{\texttt{#1}}\fi
\expandafter\ifx\csname urlprefix\endcsname\relax\def\urlprefix{URL }\fi
\providecommand{\bibinfo}[2]{#2}
\providecommand{\eprint}[2][]{\url{#2}}

\bibitem[{\citenamefont{Choptuik}(1993)}]{choptuik93}
\bibinfo{author}{\bibfnamefont{M.~W.} \bibnamefont{Choptuik}},
  \bibinfo{journal}{Phys. Rev. Lett.} \textbf{\bibinfo{volume}{70}},
  \bibinfo{pages}{9} (\bibinfo{year}{1993}).

\bibitem[{\citenamefont{Brugmann}(1996)}]{brugmann}
\bibinfo{author}{\bibfnamefont{B.}~\bibnamefont{Brugmann}},
  \bibinfo{journal}{Phys. Rev.} \textbf{\bibinfo{volume}{D54}},
  \bibinfo{pages}{7361} (\bibinfo{year}{1996}), \eprint{gr-qc/9608050}.

\bibitem[{\citenamefont{Imbiriba et~al.}(2004)}]{centrella2}
\bibinfo{author}{\bibfnamefont{B.}~\bibnamefont{Imbiriba}} \bibnamefont{et~al.}
  (\bibinfo{year}{2004}), \eprint{gr-qc/0403048}.

\bibitem[{\citenamefont{Papadopoulos et~al.}(1998)\citenamefont{Papadopoulos,
  Seidel, and Wild}}]{wild}
\bibinfo{author}{\bibfnamefont{P.}~\bibnamefont{Papadopoulos}},
  \bibinfo{author}{\bibfnamefont{E.}~\bibnamefont{Seidel}}, \bibnamefont{and}
  \bibinfo{author}{\bibfnamefont{L.}~\bibnamefont{Wild}},
  \bibinfo{journal}{Phys. Rev.} \textbf{\bibinfo{volume}{D58}},
  \bibinfo{pages}{084002} (\bibinfo{year}{1998}), \eprint{gr-qc/9802069}.

\bibitem[{\citenamefont{Diener et~al.}(2000)\citenamefont{Diener, Jansen,
  Khokhlov, and Novikov}}]{diener}
\bibinfo{author}{\bibfnamefont{P.}~\bibnamefont{Diener}},
  \bibinfo{author}{\bibfnamefont{N.}~\bibnamefont{Jansen}},
  \bibinfo{author}{\bibfnamefont{A.}~\bibnamefont{Khokhlov}}, \bibnamefont{and}
  \bibinfo{author}{\bibfnamefont{I.}~\bibnamefont{Novikov}},
  \bibinfo{journal}{Class. Quant. Grav.} \textbf{\bibinfo{volume}{17}},
  \bibinfo{pages}{435} (\bibinfo{year}{2000}), \eprint{gr-qc/9905079}.

\bibitem[{\citenamefont{New et~al.}(2000)}]{centrella}
\bibinfo{author}{\bibfnamefont{K.~C.~B.} \bibnamefont{New}}
  \bibnamefont{et~al.}, \bibinfo{journal}{Phys. Rev.}
  \textbf{\bibinfo{volume}{D62}}, \bibinfo{pages}{084039}
  (\bibinfo{year}{2000}), \eprint{gr-qc/0007045}.

\bibitem[{\citenamefont{Hern}(1999)}]{hernthesis}
\bibinfo{author}{\bibfnamefont{S.~D.} \bibnamefont{Hern}}, Ph.D. thesis,
  \bibinfo{school}{Cambridge University} (\bibinfo{year}{1999}),
  \eprint[http://arXiv.org/abs]{gr-qc/0004036}.

\bibitem[{\citenamefont{Pretorius and Lehner}(2003)}]{pretorius03}
\bibinfo{author}{\bibfnamefont{F.}~\bibnamefont{Pretorius}} \bibnamefont{and}
  \bibinfo{author}{\bibfnamefont{L.}~\bibnamefont{Lehner}}
  (\bibinfo{year}{2003}), \eprint{gr-qc/0302003}.

\bibitem[{\citenamefont{Schnetter et~al.}(2003)\citenamefont{Schnetter, Hawley,
  and Hawke}}]{hawley}
\bibinfo{author}{\bibfnamefont{E.}~\bibnamefont{Schnetter}},
  \bibinfo{author}{\bibfnamefont{S.~H.} \bibnamefont{Hawley}},
  \bibnamefont{and} \bibinfo{author}{\bibfnamefont{I.}~\bibnamefont{Hawke}}
  (\bibinfo{year}{2003}), \eprint{gr-qc/0310042}.

\bibitem[{\citenamefont{Bruegmann et~al.}(2003)\citenamefont{Bruegmann, Tichy,
  and Jansen}}]{tichy}
\bibinfo{author}{\bibfnamefont{B.}~\bibnamefont{Bruegmann}},
  \bibinfo{author}{\bibfnamefont{W.}~\bibnamefont{Tichy}}, \bibnamefont{and}
  \bibinfo{author}{\bibfnamefont{N.}~\bibnamefont{Jansen}}
  (\bibinfo{year}{2003}), \eprint{gr-qc/0312112}.

\bibitem[{\citenamefont{Choptuik
  et~al.}(2003{\natexlab{a}})\citenamefont{Choptuik, Hirschmann, Liebling, and
  Pretorius}}]{graxi1}
\bibinfo{author}{\bibfnamefont{M.~W.} \bibnamefont{Choptuik}},
  \bibinfo{author}{\bibfnamefont{E.~W.} \bibnamefont{Hirschmann}},
  \bibinfo{author}{\bibfnamefont{S.~L.} \bibnamefont{Liebling}},
  \bibnamefont{and}
  \bibinfo{author}{\bibfnamefont{F.}~\bibnamefont{Pretorius}},
  \bibinfo{journal}{Class. Quant. Grav.} \textbf{\bibinfo{volume}{20}},
  \bibinfo{pages}{1857} (\bibinfo{year}{2003}{\natexlab{a}}),
  \eprint{gr-qc/0301006}.

\bibitem[{\citenamefont{Choptuik
  et~al.}(2003{\natexlab{b}})\citenamefont{Choptuik, Hirschmann, Liebling, and
  Pretorius}}]{graxi2}
\bibinfo{author}{\bibfnamefont{M.~W.} \bibnamefont{Choptuik}},
  \bibinfo{author}{\bibfnamefont{E.~W.} \bibnamefont{Hirschmann}},
  \bibinfo{author}{\bibfnamefont{S.~L.} \bibnamefont{Liebling}},
  \bibnamefont{and} \bibinfo{author}{\bibfnamefont{F.}~\bibnamefont{Pretorius}}
  (\bibinfo{year}{2003}{\natexlab{b}}), \eprint{gr-qc/0305003}.

\bibitem[{\citenamefont{Liebling}(2002)}]{myamr}
\bibinfo{author}{\bibfnamefont{S.~L.} \bibnamefont{Liebling}},
  \bibinfo{journal}{Phys. Rev.} \textbf{\bibinfo{volume}{D66}},
  \bibinfo{pages}{041703} (\bibinfo{year}{2002}), \eprint{gr-qc/0202093}.

\bibitem[{\citenamefont{Ryder}(1996)}]{ryder}
\bibinfo{author}{\bibfnamefont{L.~H.} \bibnamefont{Ryder}},
  \emph{\bibinfo{title}{Quantum Field Theory}} (\bibinfo{publisher}{Cambridge
  University Press}, \bibinfo{year}{1996}).

\bibitem[{\citenamefont{Berger and Oliger}(1984)}]{berger}
\bibinfo{author}{\bibfnamefont{M.~J.} \bibnamefont{Berger}} \bibnamefont{and}
  \bibinfo{author}{\bibfnamefont{J.}~\bibnamefont{Oliger}},
  \bibinfo{journal}{J. Comp. Phys.} \textbf{\bibinfo{volume}{53}},
  \bibinfo{pages}{484} (\bibinfo{year}{1984}).

\bibitem[{\citenamefont{Liebling et~al.}(2000)\citenamefont{Liebling,
  Hirschmann, and Isenberg}}]{Liebling:1999nn}
\bibinfo{author}{\bibfnamefont{S.~L.} \bibnamefont{Liebling}},
  \bibinfo{author}{\bibfnamefont{E.~W.} \bibnamefont{Hirschmann}},
  \bibnamefont{and} \bibinfo{author}{\bibfnamefont{J.}~\bibnamefont{Isenberg}},
  \bibinfo{journal}{J. Math. Phys.} \textbf{\bibinfo{volume}{41}},
  \bibinfo{pages}{5691} (\bibinfo{year}{2000}),
  \eprint[http://arXiv.org/abs]{math-ph/9911020}.

\end{thebibliography}

\onecolumngrid

\end{document}